\documentstyle[aps, preprint]{revtex}
\begin{document}
\tightenlines
\draft
\title{Back-reaction effects in power-law inflation}
\author{Mauricio Bellini\footnote{E-mail address: mbellini@mdp.edu.ar}}
\address{Consejo Nacional de Investigaciones Cient\'{\i}ficas y
T\'ecnicas (CONICET)\\
and\\
Departamento de F\'{\i}sica,
Facultad de Ciencias Exactas y Naturales,
Universidad Nacional de Mar del Plata, \\
Funes 3350, (7600) Mar del Plata, Buenos Aires, Argentina.}
\maketitle
\begin{abstract}
I consider a power-law inflationary model taking into account
back-reaction effects.
The interesting result is that the spectrum for the scalar
field fluctuations does not depends on the expansion rate
of the universe $p$ and that it result to be
scale invariant for cosmological scales.
However, the amplitude for these fluctuations depends on $p$.
\end{abstract}
\vskip .2cm                             
\noindent
Pacs numbers: 98.80.Cq\\
\vskip 1cm
An inflationary phase eliminates the major problems in cosmology, namely
flatness, the horizon problem, homogeneity and the numerical density
of monopoles\cite{1,1a,1b,1c}. However, the main problem is how to attach
the observed universe to the end of the inflationary epoch.
In the standard (isentropic) inflationary scenarios, the radiation
energy density $\rho_r$ scales with the inverse fourth power
of the scale factor, becoming quickly negligible. In such case, a
short time reheating period terminates the inflationary period
and initiates the radiation dominated epoch.
The most widely accepted approach assumes that the
inflationary phase is driven by a quantum scalar field $\varphi $
related with a scalar potential $V(\varphi )$. Within this
perspective, stochastic inflation proposes to describe the
dynamics of this quantum field on the basis of two pieces: the
spatially homogeneous and inhomogeneous components $\phi_c(t)$ and
$\phi(\vec x,t)$.
During inflation vacuum fluctuations on scales smaller than
the Hubble radius are magnified into classical perturbations
in the scalar fields on scales larger than the Hubble radius.
Classical perturbations in the scalar fields can then
change the number of $e$ folds of expansion and so lead
to classical curvature and density perturbations after inflation.
Density perturbations are thught to be responsible for the
formation of galaxies and the large scale structure of the observable
universe as well as, in combination with the gravitational waves produced
during inflation, for the anisotropies in the cosmic microwave background.
The study of back-reaction effects in a single field model of inflation
is the main subject of interest in this note.

I shall consider a scalar field $\varphi(\vec x,t)$ which can be
decomposed by the semiclassical manner $\varphi(\vec x,t) = \phi_c(t) +
\phi(\vec x,t)$, such that $\left<\phi(\vec x,t)\right>=0$,
$\left<\dot\phi(\vec x,t)\right>=0$ and
$\left<\varphi(\vec x,t)\right>=\phi_c(t)$ on a background spatially
flat, Friedmann-Robertson-Walker (FRW) metric.
It is well known that fluctuations $\phi$ induce scalar metric
fluctuations $\Phi$ such that
$ds^2  = -(1+2\Phi)dt^2 + (1-2\Phi^2)a^2(t) dr^2$\cite{2-,2,3,4},
but the aim of this work is the study 
of the Hubble inhomogeneities effects
$\delta H=H'_c(\phi_c) \phi(\vec x,t)$, on the dynamics of the
$k$-modes of the $\phi$-fluctuations.

The dynamics of $\phi_c$ during inflation is given by
\begin{equation}
\ddot\phi_c + 3 H_c \dot\phi_c + V'(\phi_c) = 0,
\end{equation}
where the prime denotes the derivative with respect to the field.
On the other hand, if we expand the Hubble parameter
\begin{equation}
H(\varphi) = H_c(\phi_c) + H'(\phi_c) \phi,
\end{equation}
the Friedmann equation that describes the dynamics of the Hubble parameter
in terms of the potential $V$ will be
\begin{equation}\label{2}
\frac{3}{8\pi G}\left< H^2 \right> =
\left< \frac{\dot\varphi^2}{2} +\frac{1}{2a^2} \left(
\nabla \varphi\right)^2 + V(\varphi)\right> ,
\end{equation}
which can be expanded as
\begin{equation}\label{22}
\frac{3}{8\pi G}\left< H^2 \right> =
\left[ \frac{\dot\phi^2_c}{2} + V(\phi_c) +
\frac{\left<\dot\phi^2\right>}{2}+\frac{1}{2a^2} \left<\left(
\nabla \phi\right)^2\right> + \frac{V''(\phi_c)}{2}
\left<\phi^2\right>\right],
\end{equation}
on the background spatially flat FRW
metric $ds^2 = -dt^2 + a^2(t) dr^2$. Here, the scale factor evolves
as $a \sim e^{\int <H> dt} \equiv e^{\int H_c dt}$, such that
$H_c = \dot a/a$. However, the effective scale factor $a_{eff}$ in eq.
(\ref{22}) due to squared $\phi$-fluctuations, would be
$a_{eff} = a^{(0)}_{eff} \  e^{\int \sqrt{<H^2>} dt}$, where
$a^{(0)}_{eff}$ is the scale factor when inflation begins. Such that
effect would change the background metric $ds^2 = -dt^2 +a^2 dr^2
\rightarrow ds^2 = - dt^2 + a^2_{eff} dr^2$.
Note that eq. (\ref{22}) give us the expectation value for energy density.

If we make a Fourier expansion for the redefined fluctuations
$\chi=a^{3/2}\phi(\vec x,t)$
\begin{equation}
\chi(\vec x,t) = \frac{1}{(2\pi)^{3/2}} {\Large\int} d^3k \left[a_k \xi_k(t)
e^{i\vec k.\vec x} + h.c.\right],
\end{equation}
the dynamics of the k-modes may be affected due to actually, the
dynamics of $\xi_k$ is described by
\begin{equation}\label{4}
\ddot\xi_k + a^{-2} \left[ k^2 - k^2_0\right]\xi_k(t)=0,
\end{equation}
where
\begin{equation}\label{5}
k^2_0 = a^2 \left[\frac{9}{4} H^2_c+ \frac{3}{2} \dot H_c - V''(\phi_c)
-3 H'_c(\phi_c) \dot\phi_c\right].
\end{equation}
Note that the last term in eq. (\ref{5}) appears as a consequence of the
$\phi$-first order expansion of the Hubble parameter.
Using $\dot\phi_c(t) = -{1 \over 4\pi G} H'_c$ and $H'_c
=\dot H_c /\dot\phi_c$, the equation (\ref{5}) can be written as
\begin{equation}\label{6}
k^2_0 = a^2 \left[\frac{9}{4} H^2_c- \frac{3}{2} \dot H_c - V''(\phi_c)
\right].
\end{equation}
The wavenumber $k_0(t)$ separates the ultraviolet (UV) and infrared (IR)
sectors, where the wavenumbers $k$ are respectively
$k \gg k_0$ and $k \ll k_0$.
The condition to inflation takes place is now $p>4/3$.
The equation (\ref{2}) without consider the inhomogeneities
for $H$ and $V(\varphi)$, gives
\begin{equation}\label{7}
H^2_c =
\frac{8\pi G}{3} \left[\frac{\dot\phi^2_c}{2}
+ V(\phi_c)\right] .
\end{equation}
In power-law inflation the scale factor evolves as
$a \sim t^p$ and the Hubble parameter $H_c = \dot a/a = p/t$.
Furthermore, the
equation for the modes (\ref{4}) it is assumed a time dependent
wavenumber $k_0(t)$
\begin{equation}\label{11}
\frac{k^2_0(t)}{a^2} = t^{-2} \left[\frac{9}{4} p^2 -\frac{9}{2}p+2\right],
\end{equation}
where now appears the term $-9p/2 $ instead of $-15p/2$
inside the
brackets [see refs.\cite{bcms,hab}], due
to the additional term $3H'_c\dot\phi_c \phi$ in the equation
for the fluctuations. Hence, the fluctuations of the Hubble parameter
$H'_c\phi$ modify the spectrum of the fluctuations $\phi$.
The general solution for the eq. (\ref{4}) with time dependent wavenumber
$k_0(t)$ is
\begin{equation}\label{111}
\xi_k(t) = A_1 \sqrt{t} {\cal H}^{(1)}_{\nu}\left[
\frac{k t^{1-p} H_0}{(p-1) t^{-p}_0}\right]+ A_2
\sqrt{t} {\cal H}^{(2)}_{\nu}\left[
\frac{k t^{1-p} H_0}{(p-1) t^{-p}_0}\right],
\end{equation}
where ${\cal H}^{(1,2)}[x(t)]={\cal J}_{\nu}[x(t)]\pm {\em i} {\cal Y}_{\nu}[x(t)]$
are the Hankel functions with $\nu ={3\sqrt{(p-1)^2} \over 2(p-1)}=\pm 3/2$
for $p >1$ and $p<1$, respectively and
$x(t) = \frac{k t^{1-p} H_0}{(p-1) t^{-p}_0}$ is a dimensionless function.

Note that the value for $\nu$ is not a function of $p$ during inflation.
This is an important difference with the case of power-law inflation
when the term $3 H'_c\dot\phi$
is not considered in eq. (\ref{4}). In that case\cite{bcms},
$\nu ={\sqrt{9/4 p^2-15/2 p +9/4} \over (p-1)}$
and the range of possible $\nu$-values during inflation is: $
\nu_{p \rightarrow -\infty} \rightarrow -\infty < v
< \nu_{p \rightarrow \infty}
\rightarrow \infty$, where $\nu$ is not defined in the range
$1/3 < \nu < 3$.

Since in our case $\nu = 3/2$, the
solution for eq. (\ref{4}) is 
\begin{equation}\label{10''}
\xi_k(t) = - \sqrt{\frac{1}{2(p-1)}} \left\{
\frac{x(t) \  \cos[x(t)] - \sin[x(t)] + i \  \left[\cos[x(t)]+
x(t) \  \sin[x(t)]\right]}{x^{3/2}(t)}\right\}.
\end{equation}
Hence, the 
squared fluctuations $\left<\chi^2\right>$ 
will be
\begin{equation}\label{c} 
\left<\chi^2\right> =
\frac{t}{4\pi^2(p-1)} \left[ \left(\frac{(p-1) t^{p-1}}{t^p_0 H_o}\right)^3
{\Large \int}_{k_{min}}^{k_{p}} \frac{dk}{k} 
+ \left(\frac{(p-1) t^{p-1}}{t^p_0 H_o}\right)
{\Large \int}_{k_{min}}^{k_{p}} dk \  k \right], 
\end{equation}
where $k_{p}=1$ (in Planckian unities)
is  the Planckian cut-off  and $k_{min}=a_0/a(t) \ll 1$ for $t\gg t_0$.
Note that the second term in eq. (\ref{c}) becomes zero at the
end of power-law inflation (i.e., for $p=1$).
The first integrate dominates on very small scales and the
second one on very large scales.
The original squared fluctuations are recovered by means of the
map $ \left<\phi^2\right> = a^{-3}\left<\chi^2\right>$.

On the other hand the expectation value for the
squared gradient is
\begin{equation}\label{e}
\left<\left(\nabla\chi\right)^2\right> =
-\frac{t}{4\pi^2 (p-1)}   \left[ \left(\frac{(p-1) t^{p-1}}{t^p_0 H_o}\right)^3
{\Large \int}_{k_{min}}^{k_{p}} dk \  k +
\left(\frac{(p-1) t^{p-1}}{t^p_0 H_o}\right)
{\Large \int}_{k_{min}}^{k_{p}} dk \  k^3\right], 
\end{equation}
where $ \left<\left(\nabla\phi\right)^2\right> = a^{-3}\left<
\left(\nabla\chi\right)^2\right>$, which rapidly becomes small 
during inflation.

Since $\phi = a^{-3/2} \chi$, the squared $\dot\phi$-fluctuations can be
written as
\begin{equation}\label{h}
\left<\dot\phi^2\right> =
\frac{9}{4} a^{-3} H^2_c \left<\chi^2\right> + a^{-3} \left<\dot\chi^2\right>
-\frac{3}{2} a^{-3} H_c \left[\left<\chi\dot\chi\right> + \left<\dot\chi \chi
\right>\right],
\end{equation}
where the squared $\dot\chi$-fluctuations and the
($\chi\dot\chi + \dot\chi \chi$)-correlations are
\begin{eqnarray}
&& \left<\dot\chi^2\right>=
\frac{1}{2\pi^2} \left[\left(\frac{t}{t_0}\right)^p \frac{1}{H_0}
\right] t^{-4} \left(\frac{9}{8} p^4 - \frac{15}{4} p^3 + \frac{37}{8}
p^2 - \frac{5}{2} p + \frac{1}{2} \right)
{\Large\int}_{k_{min}}^{k_{p}} \frac{dk}{k} \nonumber \\
&&-\frac{1}{2\pi^2} \left[\left(\frac{t}{t_0}\right)^p \frac{1}{H_0}
\right] t^{-2} \left(\frac{3}{8} p^2 - p \right)
{\Large\int}_{k_{min}}^{k_{p}}
dk \  k + \frac{1}{4\pi^2} \left[H_0 \left(\frac{t_0}{t}\right)^p\right]
{\Large\int}_{k_{min}}^{k_{p}} dk \  k^3, \label{ccc} \\
&& \left<\chi \dot\chi + \dot\chi \chi\right> = \frac{1}{2\pi^2}
\left[ \left(\frac{t}{t_0}\right)^p \frac{t^{-1}}{H_0} \right]^3
\left(\frac{3}{2} p^3 - 4p^2 + \frac{7}{2} p -1\right)
{\Large\int}_{k_{min}}^{k_{p}} \frac{dk}{k}
+ \frac{1}{4\pi^2} \left[\left(\frac{t}{t_0}
\right)^p \frac{t^{-1}}{H_0} \right]
{\Large\int}_{k_{min}}^{k_{p}} dk \  k. \label{ddd}
\end{eqnarray}
Note that the first terms in eqs. (\ref{ccc}) and (\ref{ddd}) are
zero for $p=2/3, 1$ [see eq. (\ref{22})].
The expectation value for radiation energy density $<\rho>=
{3\over 8\pi G} <H^2>$ will be
\begin{equation}\label{uuu}
\frac{3}{8\pi G} \left< H^2 \right> =
\rho(\phi_c,\dot\phi_c) + \frac{1}{2 a^3} \left[ \left<\chi^2\right> \left[
\frac{9}{4} H^2_c + V''(\phi_c)\right] + \left<\dot\chi^2\right> -
\frac{3}{2} \left[\left<\chi \dot\chi + \dot\chi \chi\right>\right]
+ \frac{1}{a^2} \left<\left(\nabla\chi\right)^2\right>\right],
\end{equation}
where $\left<\chi^2\right>$, $\left<(\nabla\chi)^2\right>$,
$\left<\dot\chi^2\right>$ and $\left<\chi\dot\chi
+ \dot\chi \chi \right>$ are given respectively by eqs.
(\ref{c}), (\ref{e}), (\ref{ccc}), (\ref{ddd}) and
\begin{equation}
\rho(\phi_c,\dot\phi_c) = \frac{\dot\phi^2_c}{2} + V(\phi_c).
\end{equation}
It is easy to demonstrate
that the second term in eq. (\ref{uuu}) is very small at the
end of power-law inflation. This term describes the back-reaction
contribution for $<\rho >$ and should be the main contribution
for the cosmological constant at the end of inflation.

To summarize, in this work I studied the spectrum for scalar fluctuations
in power-law inflation taking into account the first order fluctuations
of the Hubble parameter. It changes the equation for the $\xi_k$-modes of the
redefined quantum field $\chi$, which now becomes
\begin{displaymath}
\ddot\xi_k + a^{-2} \left[ k^2 - a^2 \left(\frac{9}{4} H^2_c
-\frac{3}{2} \dot H_c - V''(\phi_c)\right)\right] \xi_k =0,
\end{displaymath}
rather than
\begin{displaymath}
\ddot\xi_k + a^{-2} \left[ k^2 - a^2 \left(\frac{9}{4} H^2_c
+\frac{3}{2} \dot H_c - V''(\phi_c)\right)\right] \xi_k =0.
\end{displaymath}
The interesting of the result for the new equation is that $\nu = 3/2$
[see eq. (\ref{111})] does not depends on the value of $p$ [for a
scale factor that evolves as: $a \sim t^p$], and so
the same power spectrum
of the $\phi$-squared fluctuations result to be the same
for any scale factor
rate expansion $p$. For super Hubble scales, the spectrums for
$\left<\chi^2 \right>$, $\left<\dot\chi^2 \right>$
and $\left< \chi\dot\chi + \dot\chi \chi \right>$ are scale invariant.
However, the amplitude for the fluctuations depends 
on $p$. On the other hand, at the end of inflation the correlations
$\left<\chi\dot\chi + \dot\chi \chi \right>$ are frozen and
($\left<\phi^2\right>$, $\left< \dot\phi^2 \right>$)
become very small.\\

\vskip .2cm
\centerline{{\bf ACKNOWLEDGEMENTS}}
\vskip .1cm
\noindent
I would like to acknowledge CONICET and AGENCIA and Universidad
Nacional de Mar del Plata (Argentina) for support.\\

\end{document}